# Magnetic Flux Diffusion and Expulsion with Thin Conducting Sheets

Yao Liu[a)]
Department of Physics, Columbia University
New York, New York 10027

John W. Belcher
Department of Physics, Massachusetts Institute of Technology
Cambridge, Massachusetts 02139

**Abstract**

We present visualizations of the diffusion and expulsion of magnetic flux for thin conducting sheets, both stationary and moving, including representations of the eddy currents and of the associated magnetic fields. Such visualizations can play an important role in making the abstract mathematics of eddy current phenomena more understandable from a physical and conceptual point of view.

## I. INTRODUCTION

Every physics student learns about eddy currents in an introductory course in electromagnetism, and the effects of such currents are easy to demonstrate. Although a straightforward application of Lenz's Law gives the sense of the circulation of eddy currents in simple situations, the analytic expressions for the currents themselves are notoriously hard to obtain even in the simplest geometries[1,2,3]. Even experienced physics instructors sometimes do not have intuition about the form that eddy currents take in more complicated situations.

In the present paper we revisit the classic papers by Saslow[4,5], who revived Maxwell's "receding image method" for calculating eddy currents for moving magnetic monopoles. Although we derive two new results here, our emphasis is on the visualization of the solutions given in Saslow. We direct the readers to Saslow for a careful derivation and a discussion of the limitations of the receding image method. Our notation follows Saslow for easy reference, and some of his arguments and formulae are reproduced here for completeness.

This paper is divided into the following sections. In Section II we consider the time evolution of the magnetic field of a monopole which suddenly appears above a thin conducting sheet. This is the basic building block for all subsequent visualizations we consider. In Appendix A, we consider the related case of a monopole appearing near two perpendicular conducting sheets. This is a new result. In Section III, we consider a monopole moving parallel to a thin conducting sheet at constant speed. The formulae for the magnetic potential and the eddy currents are given in full, with two limiting cases considered. In principle, any other magnetic configuration can be viewed as a superposition of moving/stationary magnetic monopoles, as Maxwell originally observed. However, the case of the magnetic dipole is of particular relevance physically, and involves some subtleties. In Section IV, we consider a dipole moving parallel to the sheet with a dipole moment vector parallel to the sheet and perpendicular to the direction of motion. In Appendix B, we derive the torque on a dipole moving parallel to the sheet with a magnetic dipole moment pointing in an arbitrary direction. This is a new result.

## II. STATIONARY MAGNETIC MONOPOLE

Consider the problem of a stationary magnetic monopole $q^*$ spontaneously appearing at $t = 0$ at $(x,y,z) = (0,0,h)$ above a thin conducting sheet with thickness $d$ and conductivity $\sigma$, located at $z = 0$ (see Fig. 1). At $t = 0$, when the monopole appears, there is no magnetic field for $z < 0$. As time progresses the field diffuses through the conducting sheet into the region $z < 0$. At $t = \infty$ this process is complete, and we have the field of a free monopole everywhere in space.

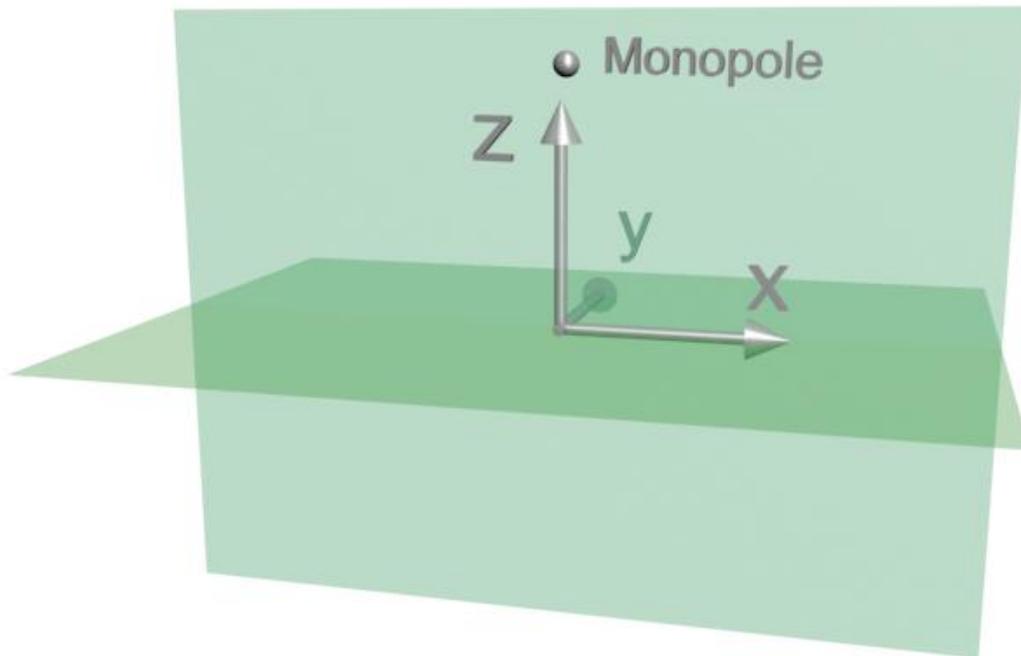

**Fig. 1:** The monopole appears at a distance $h$ above the $z = 0$ plane.

According to Maxwell's method of receding images, the magnetic field above the sheet for $t > 0$ is the same as if the sheet were not present but rather a magnetic monopole of the same charge $q^*$ were placed at $(0,0,-h)$ at $t = 0$. As time progresses this image charge recedes to $z = -\infty$ along the negative $z$-axis at constant speed

$$v_0 = \frac{2}{\mu_0 \sigma d} \ . \tag{1}$$

Thus the characteristic time scale for the magnetic field to change configuration significantly after $t = 0$ is given by

$$t_o = 2h/v_0 = \mu_0 \sigma d h. \qquad (2)$$

Similarly, the magnetic field below the sheet for $t > 0$ is as if an image monopole with magnetic charge $-q^*$ were placed at $(0,0,h)$ at $t = 0$. As time progresses the imaginary monopole recedes upward at the same speed given in equation (1). Note that this moving image charge is in addition to our original charge at $(0,0,h)$. It follows from this solution that there is a discontinuity in **B** across the plane $z = 0$ given by

$$\Delta \mathbf{B} = \frac{\mu_0 q^*}{2\pi} \frac{x\hat{\mathbf{x}} + y\hat{\mathbf{y}}}{\left[x^2 + y^2 + (h+v_0 t)^2\right]^{3/2}}. \qquad (3)$$

This discontinuity in field implies that there is an eddy current in the sheet given by

$$\mathbf{K} = \frac{1}{\mu_0}(\hat{z} \times \Delta \mathbf{B}) = \frac{q^*}{2\pi} \frac{-y\hat{\mathbf{x}} + x\hat{\mathbf{y}}}{\left[x^2 + y^2 + (h+v_0 t)^2\right]^{3/2}}. \qquad (4)$$

The magnitude of this circular current is zero at the origin and at large distances, and attains its maximum at radius $\rho = \sqrt{x^2 + y^2} = (h+v_0 t)/\sqrt{2}$. Also note that $\nabla \cdot \mathbf{K} = 0$, and thus no charge accumulates in the conducting sheet. The total current across a given horizontal direction is given by

$$I = \int_0^\infty K(\rho)\,d\rho = \frac{q^*}{2\pi} \frac{1}{h+v_0 t} \qquad (5)$$

and the total energy dissipated in the interval $0 \leq t \leq \infty$ is

$$Q = \int_0^\infty dt \int_{-\infty}^\infty dx \int_{-\infty}^\infty dy \frac{K^2}{\sigma d} = \frac{2\pi}{\sigma d}\left[\frac{q^*}{2\pi}\right]^2 \frac{1}{4v_0 h} = \frac{1}{2}\frac{\mu_0}{4\pi}\frac{q^{*2}}{2h}. \qquad (6)$$

This last expression shows that the choice of $v_0$ in Equation (1) is consistent with the conservation of energy. Note that when the sheet is perfectly conducting ($\sigma = \infty$), the quantity $v_0$ is zero, the eddy current does not decay, and the region below the sheet is permanently shielded from the field of the monopole above the sheet.

Fig. 2 shows the field configuration for three different times: (a) just after the monopole appears, at $t = 0$; (b) at a later time, $t = 0.06\,t_o$; and (c) at a still later time, $t = 0.40\,t_o$. The field configuration is shown using both field lines as well as the line integral convolution (LIC) method[6]. The LIC method has the advantage that we can code the grayscale or color of the representation according to the local strength of the magnetic field. In Fig. 2, black represents zero field and white represents high field.

As time evolves, an animation of a sequence of images like these shows the magnetic field strength in the region *z < 0* evolve from zero to the configuration expected for a free monopole. We note that this type of solution can be extended to the case of a monopole appearing near two perpendicular thin sheets, and we give the details of that solution in Appendix A and show animations of it in [7].

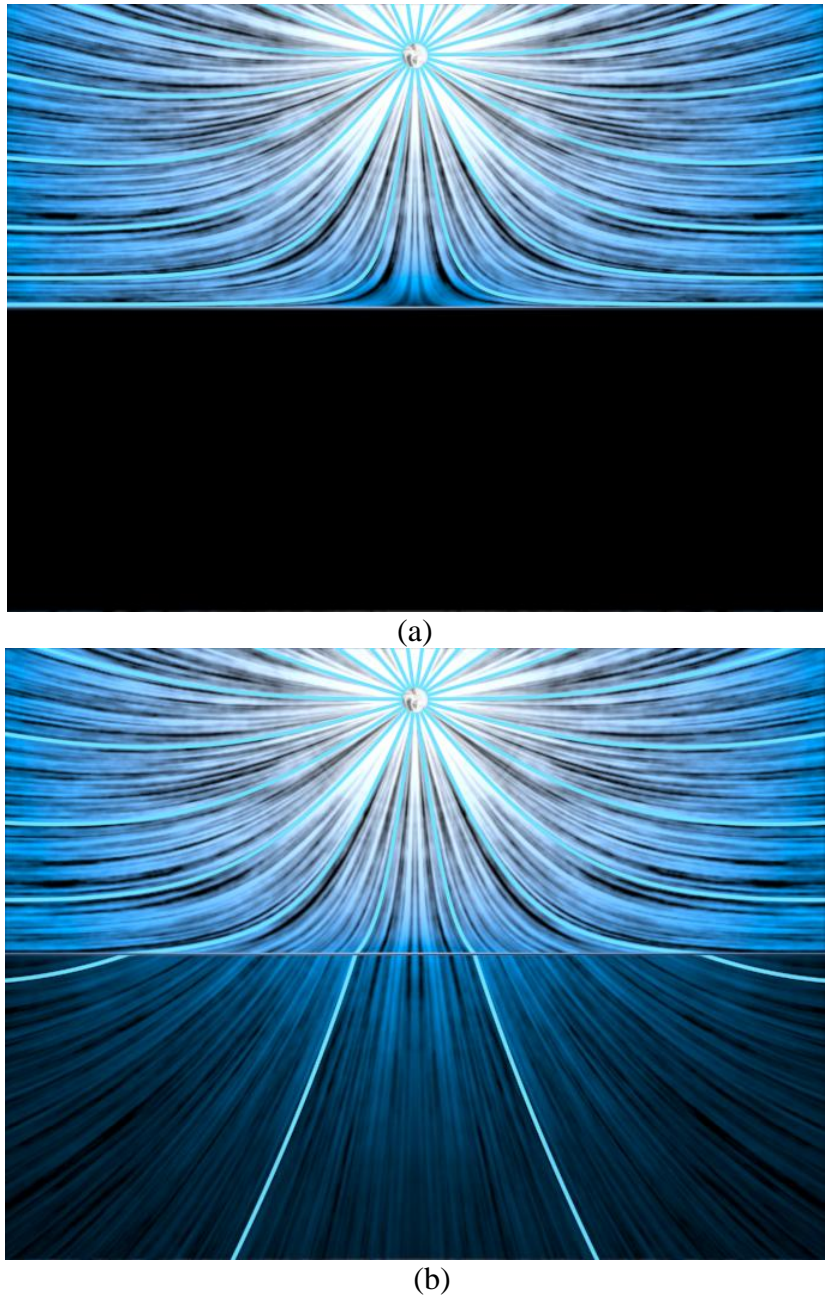

(a)

(b)

**Fig. 2:** The field of the monople at two different times after it appears at *(0,0,h)* at *t = 0*. The darker regions indicate weaker magnetic field strength.

### III. MOVING MONOPOLE

The power of Maxwell's method of receding images rests in its extension to the case of a monopole moving horizontally near a thin conducting plane. Consider a monopole at $(0,0,h)$ above the origin which has come from the left with constant velocity $v\hat{x}$ (cf. Fig. 1). To understand how we construct the solution to this problem, imagine time to be broken down into discrete time intervals of length $\tau$. At each time step, we imagine that minus and plus monopoles appear at the previous and current charge positions to reproduce the progression of the charge from left to right. The corresponding image charges as discussed above also appear, and upon creation start receding at the characteristic speed $v_0$. In the region $z > 0$, the magnetic scalar potential is the potential of the monopole $q^*$ at $(0,0,h)$ plus the potential $\Phi_+(x,y,z)$ due to the receding images below the plane. These images form a trail of plus and minus charges extending to infinity with slope $v_0/v$. As we let $\tau \to 0$, these minus and plus image charges are smeared out to form a uniform line charge below the plane. The image configuration below the plane then consists of a line-dipole with dipole moment per unit length

$$m = q^* \frac{vv_0}{v^2 + v_0^2} \tag{6}$$

*plus* an additional image charge

$$q_i^* = q^* \frac{v^2}{v^2 + v_0^2} \tag{7}$$

at $(0,0,-h)$.

It is worthwhile pointing out two limiting cases for the magnitude of these image charges: $v \ll v_0$ and $v \gg v_0$. In both of these cases the line dipole moment per unit length $m = 0$, and the problem simplifies to a single image monopole directly below the moving monopole at $(0,0,-h)$. At low speed, that single image charge has charge $q_i^* = 0$, and the field is solely due to the source charge. That is, the field of the moving monopole penetrates through the conducting plane. At high speed, the single charge has charge $q_i^* = q^*$, and the field is completely expelled from $z < 0$. The eddy current in the high speed case is thus circular and is analogous to the situation of a stationary monopole above a superconductor, or to the situation of the monopole we discussed in the above section at $t = 0$ (cf. equation (4)).

In the general case, the magnetic scalar potential for $z > 0$ (not including the potential due to the original dipole) due to the arrangement of image charges described above is

$$\Phi_+ = \frac{\mu_0}{4\pi} \frac{q^* v}{[v(z+h) - v_0 x]^2 + (v^2 + v_0^2)y^2} \left[ \frac{v_0^2 x - v_0 v(z+h)}{\sqrt{v^2 + v_0^2}} - \frac{v_0(z+h)x - v(z+h)^2 - vy^2}{\sqrt{x^2 + y^2 + (z+h)^2}} \right]. \tag{8}$$

Equation (8) is the same as equation (65) of [4], except that we have generalized the expression to include the *y* dependence. To obtain the lift and the drag on the monopole due to the eddy currents in the sheet, we need only compute $-q^* \nabla \Phi_+$ at the position of the monopole. Straightforward differentiation yields the following expressions for the lift $F_L$ and drag $F_D$ on the monopole:

$$F_L = \frac{\mu_0 q^{*2}}{16\pi h^2} \left( 1 - \frac{v_0}{\sqrt{v^2 + v_0^2}} \right), \tag{9}$$

$$F_D = \frac{\mu_0 q^{*2}}{16\pi h^2} \frac{v_0}{v} \left(1 - \frac{v_0}{\sqrt{v^2 + v_0^2}}\right). \tag{10}$$

The potential $\Phi_-(x, y, z)$ below the plane is due to a set of image charges above the plane whose configuration is exactly the mirror reflection about the $z = 0$ plane of the image charges below the plane, with all the signs of the charges flipped, that is

$$\Phi_-(x, y, z) = -\Phi_+(x, y, -z). \tag{11}$$

The current distribution $\mathbf{K}$ in the conducting plane is

$$\mathbf{K} = \frac{1}{\mu_0}(\hat{\mathbf{z}} \times \Delta \mathbf{B}) = -\frac{1}{\mu_0}\left[\hat{\mathbf{z}} \times (\nabla \Phi_+ - \nabla \Phi_-)\right]\bigg|_{z=0}. \tag{12}$$

Using the relation in equation (11) in (12), we have that

$$\mathbf{K} = \frac{2}{\mu_0}\left[\frac{\partial \Phi_+}{\partial y}\hat{\mathbf{x}} - \frac{\partial \Phi_+}{\partial x}\hat{\mathbf{y}}\right]\bigg|_{z=0} =$$

$$\frac{q^* v}{2\pi}\left\{\frac{2\left[v_0(v_0 x - vh)\hat{\mathbf{y}} - (v^2 + v_0^2)y\hat{\mathbf{x}}\right]}{\left[(v^2 + v_0^2)y^2 + (v_0 x - vh)^2\right]^2}\left[\frac{v_0(v_0 x - vh)}{\sqrt{v^2 + v_0^2}} + \frac{vy^2 - (v_0 x - vh)h}{\sqrt{x^2 + y^2 + h^2}}\right]\right.$$

$$\left. + \frac{1}{(v^2 + v_0^2)y^2 + (v_0 x - vh)^2}\left[-\frac{v_0^2 \hat{\mathbf{y}}}{\sqrt{v^2 + v_0^2}} + \frac{v_0 h \hat{\mathbf{y}} + 2vy\hat{\mathbf{x}}}{\sqrt{x^2 + y^2 + h^2}} + \frac{[vy^2 - (v_0 x - vh)h](x\hat{\mathbf{y}} - y\hat{\mathbf{x}})}{(x^2 + y^2 + h^2)^{3/2}}\right]\right\}. \tag{13}$$

In the limit that $v$ is large, or more precisely, $v \gg v_0$, $v \gg v_0 x/h$, and $v \gg v_0 y/h$, this formula reduces to

$$\mathbf{K} \approx \frac{\mu_0 q^*}{2\pi} \frac{x\hat{\mathbf{y}} - y\hat{\mathbf{x}}}{\left[x^2 + y^2 + h^2\right]^{3/2}}, \tag{14}$$

which as expected gives a circular eddy current. In the other extreme, for $v \ll v_0$, $v \ll v_0 x/h$, and $v \ll v_0 y/h$,

$$\mathbf{K} \approx \frac{q^*}{2\pi} \frac{v}{v_0} \frac{1}{x^2 + y^2}\left\{\frac{(x^2 - y^2)\hat{\mathbf{y}} - 2xy\hat{\mathbf{x}}}{x^2 + y^2}\left[1 - \frac{h}{\sqrt{x^2 + y^2 + h^2}}\right] - \frac{(x^2\hat{\mathbf{y}} - xy\hat{\mathbf{x}})h}{\left[x^2 + y^2 + h^2\right]^{3/2}}\right\}. \tag{15}$$

This eddy current system as viewed from above shows two counter-rotating swirls, one counterclockwise and ahead of the moving monopole, and the other lagging behind the monopole and rotating clockwise, in agreement with a qualitative analysis using Lenz's law, as shown in Fig. 3, where we use a LIC representation for the eddy currents. As we move from the lower speed to higher speed, the whirl behind shifts further behind, and the whirl in the front slowly moves to the center position just underneath the monopole, and becomes dominant.

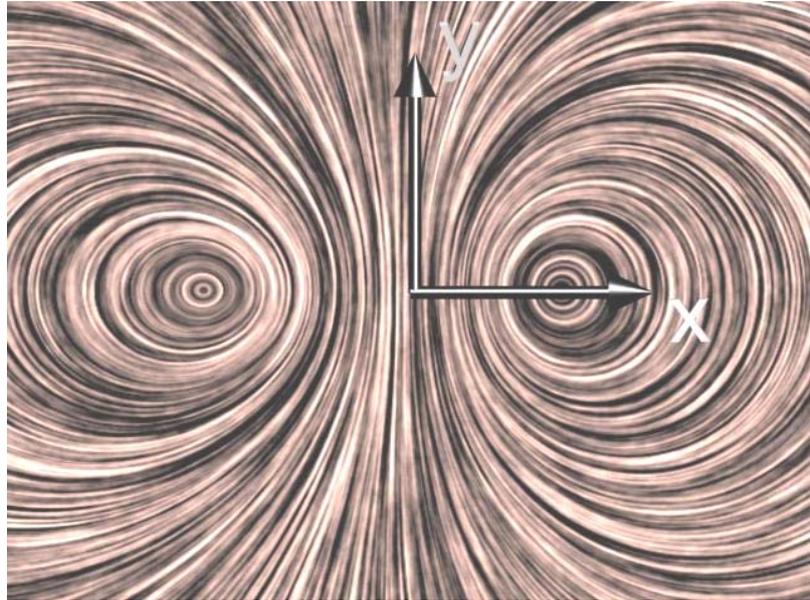

Fig. 3: The eddy current pattern for a moving monopole when $v = 0.1 v_0$. At this speed we see a slight lagging of the trailing whirl as compared to the leading whirl.

In Fig. 4, we show the field lines and the eddy current pattern in a perspective view for a moving monopole when the ratio $v/v_0$ is 3.72. Again, we use a LIC representation for the eddy currents, placed on a screen in the *xy* plane at *z = 0*, and made semi-transparent so we can see the field lines beneath the conducting sheet. If the monopole were not moving, the field lines shown would lie on a cone of half angle 30 degrees from the *–z* axis. In the figure, the field lines are of course distorted from this cone by the eddy currents flowing in the sheet.

Note that the field lines in Fig. 4 are preferentially excluded from the region *z < 0* ahead of the moving monopole, and that the field lines behind the moving monopole are "hooked" back into the direction of motion once they move into the region *z < 0*. This behavior is more clearly shown in Fig. 5, which is in the same format as Fig. 2. In this figure we show a set of field lines which would be straight lines in the *yz* plane radiating from the position of the monopole and spaced 30 degrees apart if the monopole were not moving, but which are distorted because of the eddy current system in the sheet. One of the interesting things apparent in the animations is how high v has to become before all the trailing field lines are finally expelled from the region *z < 0*. For example, the "hooked" field line in the left bottom of Fig. 4 is not finally expelled from the region *z < 0* until the ratio $v/v_0$ is about 60. However, even though the field lines are not expelled the field strength has diminished significantly, and one can justifiably ignore the effects of the field long before the field is totally expelled from the region *z < 0*.

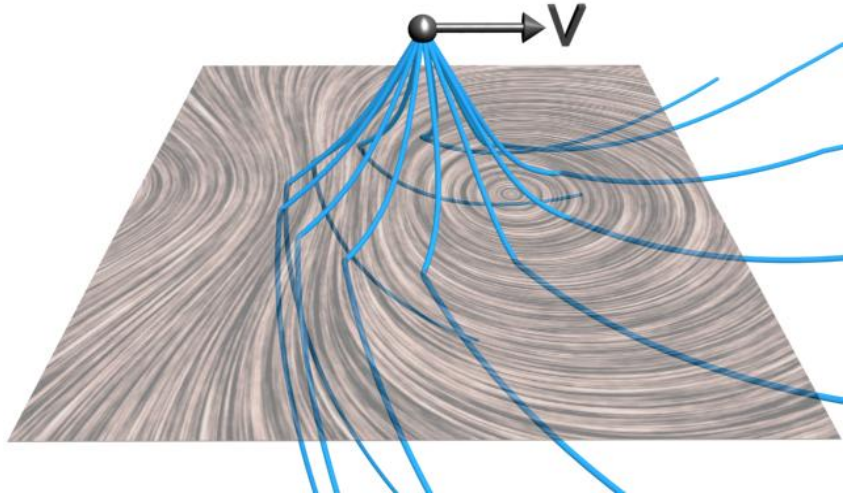

**Fig. 4:** Representative magnetic field lines and the eddy current pattern for a monopole moving with speed for v = 3.72 $v_0$.

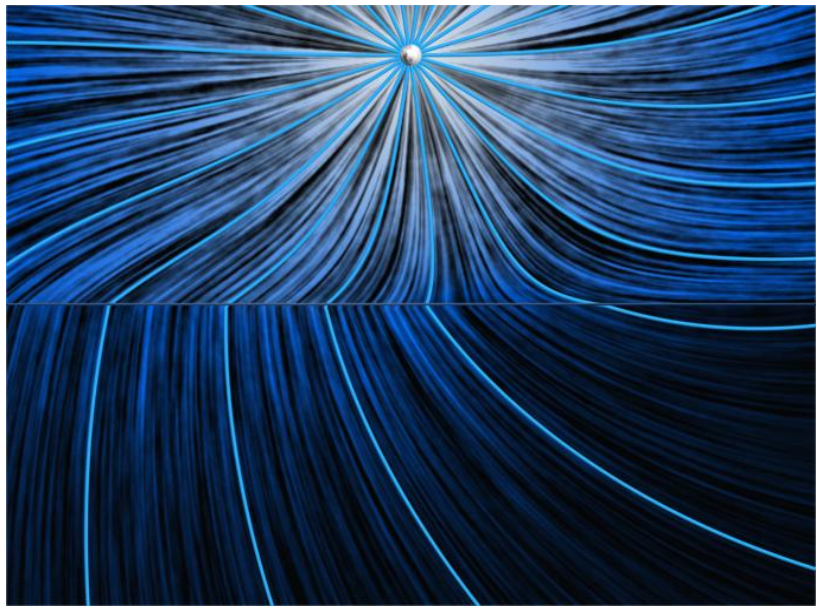

**Fig. 5:** The magnetic field configuration in the *yz* plane showing both field lines and a LIC of the magnetic field in this plane coded to indicate field strength, as in Fig. 2.

## IV. MOVING DIPOLE

Finally, we come to the most physically relevant case, that of a magnetic dipole moving parallel to a thin conducting plane. Here we consider only a dipole whose dipole moment vector lies in the *xy* plane and is oriented in the +*y* direction, that is perpendicular to the direction of motion. Other orientations are considered in Appendix B, where we also compute the torque on the moving dipole. It is shown there that if the dipole is free to rotate in the *xy* plane and only that plane, the torque on the dipole is such that orientation along the $\pm\hat{y}$ direction is a stable equilibrium, whereas orientation along the $\pm\hat{x}$ direction (that is, parallel or anti-parallel to the direction of motion) is an unstable equilibrium. This effect is easily seen using rare earth magnets above moving conducting disks.

In Fig. 6, we show the field lines and the eddy current pattern for a moving dipole where the ratio $v/v_0$ is 8.1. As in Fig. 4, we use a LIC representation for the eddy currents, placed on a screen in the *xy* plane at *z = 0*, and made semi-transparent so we can see the field lines beneath the conducting sheet. If the dipole were not moving, the field lines shown would be typical dipolar field lines, symmetric about the direction of the dipole and separated in azimuth by 30 degrees. In the figure, the field lines are distorted from this configuration by the eddy currents flowing in the sheet.

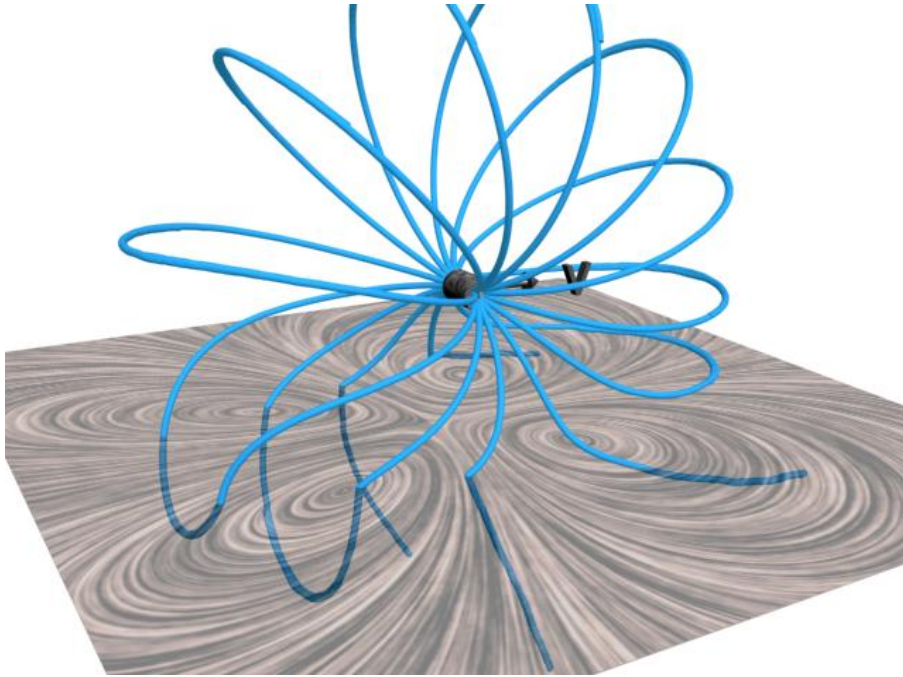

**Fig. 6:** Representative magnetic field lines and the eddy current pattern for a dipole moving with speed for v = 8.1 $v_0$. The dipole direction is in the *xy* plane in the *+y* direction.

Again, as in the moving monopole case, note the field lines are preferentially excluded from the region *z < 0* ahead of the moving dipole, and the field lines behind the moving monopole are "hooked" back into the direction of motion once they move into the region *z < 0*. The eddy current pattern now consists of four counter-rotating swirls, with adjacent swirls rotating in the opposite sense.. This pattern can be seen as a natural superposition of two eddy current patterns that we saw above in the monopole case, one for a positive monopole and one for a negative monopole, separated by a distance along the *+y* direction. As we move from lower speed to higher speed, the whirls behind shift further behind, and the whirls in the front slowly move to a position underneath the dipole, and become dominant.

In the case that the dipole moment is in the *+z* direction case, the eddy current patterns are very similar to the patterns of the monopole currents shown in Fig. 4, with similar limiting behaviors at low and high velocity. The eddy current patterns for the $+x$ case are quite different from both the $+y$ and the $+z$ case, but again can be seen as the appropriate superposition of the two eddy current patterns that we saw above in the moving monopole case, one for a positive charge and one for a negative charge, with the monopoles separated in the $+x$ direction instead of the $+y$ direction.

## V. DISCUSSION AND CONCLUSIONS

We have revisited the classic papers by Saslow[4,5] on Maxwell's receding image method for calculating the magnetic fields and eddy currents for magnetic monopoles or dipoles moving parallel to a thin conducting sheet. Although we have derived several new results here, in particular the torque on a dipole, our emphasis has been on the visualization of the solutions given in Saslow.

We have presented a variety of ways to visualize both the fields and the eddy currents in these situations. From these visualizations, we find that for both monopoles and dipoles, as the speed v increases, the magnetic flux is preferentially excluded from the region $z < 0$ as compared to the region behind the moving monopole or dipole. Simultaneously, the field lines behind the moving monopole or dipole are "hooked" back into the direction of motion once they move into the region $z < 0$. In both cases, at very high velocity, the field lines are totally excluded from the region $z < 0$, as we expect, although this limiting behavior occurs only at high multiples of the characteristic speed $v_o$.

The eddy current patterns corresponding to these field configurations show a natural evolution from what we intuitively expect at low velocity to what we expect at high velocity, when the thin sheet effectively acts as if it has zero resistance. The eddy current patterns for the moving dipole cases can be understood as a superposition of two eddy current patterns for the moving monopole case, for oppositely charged monopoles displaced in the appropriate directions.

The visualization techniques we have presented here can play a role in making the abstract mathematics of eddy current phenomena more understandable from a physical and conceptual point of view. This is true of many problems in electromagnetism, as well as in other fields of physics and mathematics. Many of these problems would benefit from the use of the powerful programs presently available for both the creation and dissemination of visualizations. The programs and scripts we have used for creating the visualizations shown here are open source, as described at http://web.mit.edu/viz/soft/.

**ACKNOWLEDGMENTS**

This work is supported by Rabi Scholars Program of Columbia College to one of us (Liu), by a grant from the Davis Education Foundation, and by NSF Grant #06185580. We are indebted to Mark Bessette of the MIT Center for Educational Computing Initiatives for technical support.

## APPENDIX A: MONOPOLE NEAR TWO CONDUCTING THIN SHEETS

An interesting combination of the stationary and moving monopole cases is the following problem (analogous to the corresponding electrostatics image problem). A monopole is spontaneously generated in the first quadrant at $(k,0,h)$ at $t=0$ near two thin conducting planes (one in the $z=0$ plane of Fig. 1 and the other in the $x=0$ plane). The general solution in the first quadrant can be obtained using three image distributions in the other quadrants. The first two are in the second and fourth quadrant, and they are charges which recede exactly as we would expect in analogy to the solution for the single sheet discussed above, except that the characteristic recession speeds are in general different: $v_1 = 2/(\mu_0 \sigma_1 d_1)$ and $v_2 = 2/(\mu_0 \sigma_2 d_2)$ in place of the $v_0$ in previous discussion, where $d_1$ ($d_2$) and $\sigma_1$ ($\sigma_2$) are characteristic of the first (second) sheet. At $t=0$, an observer in the first quadrant sees two images located at $(-k,0,h)$, $(k,0,-h)$, and a third one at $(-k,0,-h)$ serving as the image of both images. As time progresses, the two image charges recede at velocities $-v_1 \hat{x}$ and $-v_2 \hat{z}$, and thereby in moving parallel to their respective perpendicular planes, create in the fourth quadrant a trail of plus and minus charges. The image trails as viewed by the two charges coincide exactly, consisting of a line-dipole from $(-k-v_1 t,0,-h)$ to $(-k,0,-h-v_2 t)$, with dipole moment per unit length

$$m = q^* \frac{v_1 v_2}{v_1^2 + v_2^2} \qquad (A1)$$

pointing away from the origin, and two additional charges

$$q_1^* = q^* \frac{v_1^2}{v_1^2 + v_2^2} \qquad (A2)$$

$$q_2^* = q^* \frac{v_2^2}{v_1^2 + v_2^2} \qquad (A3)$$

at the endpoints $(-k-v_1 t,0,-h)$ and $(-k,0,-h-v_2 t)$, respectively. Note that $q_1^* + q_2^* = q^*$.

Physically one can imagine that the image charge initially at $(-k,0,-h)$ splits into two pieces receding at $-v_1 \hat{x}$ and $-v_2 \hat{z}$, while a line-dipole is being formed between the two. Note in particular the case when one recession speed is much larger than the other, say $v_1 \gg v_2$. Then $m=0$, $q_1^* = q^*$, and $q_2^* = 0$, corresponding to an instant diffusion into the second quadrant, followed by a gradual diffusion into the lower half as in Section II above. The magnetic scalar potential in the other three quadrants for the general case can be obtained from the one in the first by "mirror reflecting" relevant terms, *and* charge conjugation. We give in [7] visualizations of the magnetic field configuration for this case, similar in format to those in Fig. 2.

## APPENDIX B:  FORCE AND TORQUE ON A MOVING DIPOLE

Consider a magnetic dipole moving parallel to a thin conducting plane. We first take the dipole in the $+\hat{z}$ direction as consisting of two monopoles a distance $d$ apart, keeping in mind that eventually $d \to 0$ and $q^* \to \infty$ such that the product $q^*d$ remains a constant. Since the positive charge feels the potential induced by both itself and the negative charge, the sum is in fact a difference of $\Phi_+$ in equation (8) evaluated at two points separated by $\hat{z}d$. Using a Taylor expansion in $d$, we have

$$\Phi_+(\mathbf{r}) - \Phi_+(\mathbf{r} - \hat{z}d) = \frac{\partial \Phi_+}{\partial z} d + \frac{1}{2} \frac{\partial^2 \Phi_+}{\partial z^2} d^2 + \cdots \tag{B1}$$

evaluated at *(0,0,h)*.  In calculating the total force on this dipole, the first order terms for the plus and minus monopoles are equal and opposite, and it is the second order that contributes in the lift and the drag:

$$F_L = \frac{3\mu_0 (q^*d)^2}{32\pi h^4} \left(1 - \frac{v_0}{\sqrt{v^2 + v_0^2}}\right); \tag{B2}$$

$$F_D = \frac{3\mu_0 (q^*d)^2}{32\pi h^4} \frac{v_0}{v} \left(1 - \frac{v_0}{\sqrt{v^2 + v_0^2}}\right). \tag{B3}$$

which retains the relation $F_D / F_L = v_0 / v$, as for the monopole. The first order terms, however, provide a torque on the dipole, in $+\hat{y}$ direction

$$\tau = \frac{\mu_0 (q^*d)^2}{16\pi h^3} \frac{v_0}{v} \left(1 - \frac{v_0}{\sqrt{v^2 + v_0^2}}\right). \tag{B4}$$

In the more general case, consider the dipole pointing in an arbitrary direction $\hat{\mathbf{n}} = (n_x, n_y, n_z)$. For clarity, we separate the $q^*$ out of equation (9) and let

$$\Phi_+(x, y, z) = q^* \Psi(x, y, z). \tag{B5}$$

The force on the plus monopole amounts to a first-order expansion of the **B** field as in equation (B1)

$$F_x = q^{*2} \left[ \frac{\partial^2 \Psi}{\partial x \partial x}(dn_x) + \frac{\partial^2 \Psi}{\partial y \partial x}(dn_y) + \frac{\partial^2 \Psi}{\partial z \partial x}(-dn_z) \right], \tag{B6}$$

and similarly for the other components. All the second partial derivatives are evaluated at $(0,0,h)$. Since $\Psi$ is smooth and is even in $y$, $(\partial \Psi/\partial y)|_{y=0} = 0$, and hence all the mixed partials with $y$ vanish. Therefore,

$$\mathbf{F} = q^{*2}d \left( \left(\frac{\partial^2 \Psi}{\partial x^2} n_x - \frac{\partial^2 \Psi}{\partial x \partial z} n_z\right)\hat{\mathbf{x}} + \frac{\partial^2 \Psi}{\partial y^2} n_y \hat{\mathbf{y}} + \left(\frac{\partial^2 \Psi}{\partial x \partial z} n_x - \frac{\partial^2 \Psi}{\partial z^2} n_z\right)\hat{\mathbf{z}} \right), \tag{B7}$$

and the torque is

$$\boldsymbol{\tau} = d\,\hat{\mathbf{n}} \times \mathbf{F} = (q^*d)^2 \left\{ \hat{\mathbf{x}} \left[ \frac{\partial^2 \Psi}{\partial x \partial z} n_x n_y - \left( \frac{\partial^2 \Psi}{\partial y^2} + \frac{\partial^2 \Psi}{\partial z^2} \right) n_y n_z \right] \right.$$
$$+ \hat{\mathbf{y}} \left[ \left( \frac{\partial^2 \Psi}{\partial x^2} + \frac{\partial^2 \Psi}{\partial z^2} \right) n_x n_z - \frac{\partial^2 \Psi}{\partial x \partial z} (n_x^2 + n_z^2) \right], \quad (B8)$$
$$\left. + \hat{\mathbf{z}} \left[ \frac{\partial^2 \Psi}{\partial x \partial z} n_y n_z - \left( \frac{\partial^2 \Psi}{\partial x^2} - \frac{\partial^2 \Psi}{\partial y^2} \right) n_x n_y \right] \right\}$$

where

$$\frac{\partial^2 \Psi}{\partial x \partial z} = -\frac{\mu_0}{4\pi} \frac{2}{(2h)^3} \frac{v_0}{v} \left( 1 - \frac{v_0}{\sqrt{v^2 + v_0^2}} \right), \quad (B9)$$

$$\frac{\partial^2 \Psi}{\partial x^2} = -\frac{\mu_0}{4\pi} \frac{1}{(2h)^3} \left( 1 - 2\frac{v_0^2}{v^2} \left( 1 - \frac{v_0}{\sqrt{v^2 + v_0^2}} \right) \right), \quad (B10)$$

$$\frac{\partial^2 \Psi}{\partial y^2} = -\frac{\mu_0}{4\pi} \frac{1}{(2h)^3} \left( 1 + 2\frac{v_0^2}{v^2} - 2\frac{v_0}{v^2} \sqrt{v^2 + v_0^2} \right), \quad (B11)$$

$$\frac{\partial^2 \Psi}{\partial z^2} = \frac{\mu_0}{4\pi} \frac{2}{(2h)^3} \left( 1 - \frac{v_0}{\sqrt{v^2 + v_0^2}} \right). \quad (B12)$$

all evaluated at *(0,0,h)*. As a check, note that $\nabla^2 \Psi = 0$ at *(0,0,h)*, since the image potential does not see the monopoles. Also note that unlike the situation in which a dipole is immersed in an external **B** field in which the torque is simply $\tau = \mathbf{m} \times \mathbf{B}$ and tends to align the dipole parallel to **B**, the torque in this case is quadratic in the dipole moment $\mathbf{m} = q^*\mathbf{d}$ and depends on the derivatives of **B**, and has a complicated angular dependence because the "external" field is not truly external in the sense that it is induced by the dipole itself.

If the dipole is constrained to the $xz$-plane, $\hat{\mathbf{n}} = (\cos\theta, 0, \sin\theta)$, and the torque reduces to

$$\boldsymbol{\tau} = (q^*d)^2 \hat{\mathbf{y}} \left[ \left( \frac{\partial^2 \Psi}{\partial x^2} + \frac{\partial^2 \Psi}{\partial z^2} \right) \sin\theta \cos\theta - \frac{\partial^2 \Psi}{\partial x \partial z} \right], \quad (B13)$$

which yields equation (B4) as a special case. Since the second term does not depend on the angle and is always positive, the dipole tends to rotate forward—tumbling over as it is passing above the conducting plane.

If the dipole is constrained to the $xy$ plane, $\hat{\mathbf{n}} = (\cos\theta, \sin\theta, 0)$, and the torque reduces to

$$\boldsymbol{\tau} = (q^*d)^2 \left\{ \hat{\mathbf{x}} \left[ \frac{\partial^2 \Psi}{\partial x \partial z} \sin\theta \cos\theta \right] + \hat{\mathbf{y}} \left[ -\frac{\partial^2 \Psi}{\partial x \partial z} \cos^2\theta \right] + \hat{\mathbf{z}} \left[ \left( \frac{\partial^2 \Psi}{\partial y^2} - \frac{\partial^2 \Psi}{\partial x^2} \right) \sin\theta \cos\theta \right] \right\},$$
(B14)

in which only the $z$ component is relevant if we constrain the dipole to rotate only in the *xy* plane. It turns out that $\left(\frac{\partial^2}{\partial y^2} - \frac{\partial^2}{\partial x^2}\right)\Psi > 0$ for all v, although it is small for both v $\gg$ v$_0$ and v $\ll$ v$_0$. The $\sin 2\theta$ component in the z component of the torque means therefore that the dipole has a stable equilibrium if it is oriented in the $\pm\hat{y}$ direction, and an unstable equilibrium if oriented in the $\pm\hat{x}$ direction.

If one wishes to calculate the general formula for the net force exerted on the dipole, one needs to consider further terms in equation (B1) and equation (B6). The resulting expression is algebraically involved, so we do not give it here.

Using a similar analysis, we can find the discontinuity in the magnetic field across the z = 0 plane as

$$\Delta \mathbf{B}_{dip} = 2q^*d \left\{ \hat{\mathbf{x}} \left[ \frac{\partial^2 \Psi}{\partial x^2} n_x + \frac{\partial^2 \Psi}{\partial x \partial y} n_y + \frac{\partial^2 \Psi}{\partial x \partial z}(-n_z) \right] + \hat{\mathbf{y}} \left[ \frac{\partial^2 \Psi}{\partial x \partial y} n_x + \frac{\partial^2 \Psi}{\partial y^2} n_y + \frac{\partial^2 \Psi}{\partial y \partial z}(-n_z) \right] \right\},$$
(B15)

where now all partial derivatives are evaluated at $(x, y, 0)$. The eddy current for a dipole can then be obtained from $\mathbf{K} = (\hat{\mathbf{z}} \times \Delta \mathbf{B})/\mu_0$. The analytical form is not particularly instructive, so we do not give it here.